\documentclass[conference]{IEEEtran}
\IEEEoverridecommandlockouts
\usepackage[labelformat=simple]{subcaption}

\usepackage{stfloats}

\usepackage{cite}
\usepackage{amsmath,amssymb,amsfonts}
\usepackage{algorithmic}
\usepackage{algorithm}
\usepackage{graphicx}
\usepackage{textcomp}
\usepackage{xcolor}
\usepackage{subcaption}
\usepackage{flushend}
\usepackage{mathtools}
\usepackage{tikz}
\usetikzlibrary{positioning}
\usepackage[normalem]{ulem}
\usetikzlibrary{plotmarks}
\usepackage{algorithm}\usepackage{soul}\usepackage{ulem}

\floatname{algorithm}{Algorithm}
\usepackage{algorithmic}
\usepackage{physics}
\usepackage[colorlinks=true,citecolor=blue,linkcolor=blue,urlcolor=blue]{hyperref}

\usepackage{amsthm}

\def\BibTeX{{\rm B\kern-.05em{\sc i\kern-.025em b}\kern-.08em
    T\kern-.1667em\lower.7ex\hbox{E}\kern-.125emX}}
\begin{document}

\title{Enhancing the Performance of Quantum Neutral-Atom-Assisted Benders Decomposition}
    
\author{

Anna Joliot\IEEEauthorrefmark{1},
M. Yassine Naghmouchi\IEEEauthorrefmark{1},
 Wesley Coelho\IEEEauthorrefmark{1}
\\

\\
\IEEEauthorblockA{\IEEEauthorrefmark{1}PASQAL,\textit{ 24 Av. Emile Baudot, 91120 Palaiseau, France}}
}


\maketitle
\thispagestyle{plain}
\pagestyle{plain}

\begin{abstract}
This paper presents key enhancements to our previous work~\cite{naghmouchi2024mixed} on a hybrid Benders decomposition (HBD) framework for solving mixed integer linear programs (MILPs). In our approach, the master problem is reformulated as a Quadratic Unconstrained Binary Optimization (QUBO) model and solved on a neutral-atom quantum processor using automated conversion techniques. Our enhancements address three critical challenges. First, to adapt to hardware constraints, we refine the QUBO formulation by tightening the bounds of continuous variables and employing an exponential encoding method that eliminates slack variables, thereby reducing the required qubit count. Second, to improve solution quality, we propose a robust feasibility cut generation method inspired by the L-shaped approach and implement a constructive penalty tuning mechanism that replaces manual settings. Third, to accelerate convergence, we introduce a multi-cut strategy that integrates multiple high-density Benders cuts per iteration. Extensive numerical results demonstrate significant improvements compared to our previous approach: the feasibility rate increases from 68\% to 100\%, and the optimality rate rises from 52\% to 86\%. These advancements provide a solid foundation for future hybrid quantum-classical optimization solvers.
\end{abstract}

\newcommand{\yn}[1]{\textcolor{violet}{#1}}

\section{\uppercase{Introduction}}
\label{sec:introduction}

In recent years, hybrid classical–quantum approaches have gained traction in addressing NP-hard problems~\cite{ nannicini2019performance, da2023quantum}. By combining the strengths of classical solvers with emerging quantum hardware, these methods promise to tackle complex optimization tasks that are otherwise intractable on conventional computers. Mixed integer linear programming (MILP) is a powerful modeling tool for hard optimization problems used across various industries. A well-established technique for solving MILPs is Benders Decomposition (BD)~\cite{bnnobrs1962partitioning}, which splits the problem into a master problem (MP) and a subproblem (SP). In this framework, the master problem, responsible for handling integer variables, and the subproblem, managing continuous variables, are solved iteratively. With each iteration, Benders feasibility and optimality cuts derived from the dual formulation of the subproblem are progressively incorporated into the master problem to tighten the feasible region and steer the algorithm towards an optimal solution, respectively.

Recently, hybrid BD (HBD) approaches have emerged to address the computational challenges of the MP, due to its combinatorial difficulty—originating from the binary variables—and the iterative addition of Benders cuts, which progressively expands the problem's complexity. In these frameworks, the MP is reformulated as a Quadratic Unconstrained Binary Optimization (QUBO) model, which is naturally suitable for quantum computations. This is because the quadratic form of QUBO models directly corresponds to the Hamiltonian dynamics of quantum systems, enabling efficient exploration of the solution space through various quantum computing techniques such as gate-based circuits, quantum annealing, and variational algorithms. For instance, Zhao et al.~\cite{zhao2022hybrid} and Gao et al.~\cite{gao2022hybrid} have demonstrated promising results by solving the QUBO formulation on quantum annealers, while Chang et al.~\cite{chang2020hybrid} and Fan et al.~\cite{fan2022hybrid} have successfully applied HBD techniques on noisy intermediate-scale quantum processors. Franco et al.~\cite{franco2023efficient} further compared BD to Dantzig–Wolfe, noting that although BD applies to a broader class of MILPs, it often requires more qubits.

In line with these advances, our previous work~\cite{naghmouchi2024mixed} introduced a PoC (proof-of-concept) of a novel HBD approach assisted by neutral-atom quantum processors~\cite{henriet2020quantum}—a technology with interesting promise of enhanced connectivity and scalability compared to existing quantum hardware. The master problem was converted into a QUBO model via an automated procedure and solved on a neutral-atom device. This pipeline relied on two key components: \emph{register embedding}, which arranges atoms in a geometric register that mirrors the QUBO interactions, and \emph{variational pulse shaping}, which optimizes quantum control parameters (e.g., Rabi frequency, detuning, and pulse duration) with outcomes mapped to cost values through the QUBO Hamiltonian and refined using Gradient Boosted Regression Trees. Our results were promising, achieving good feasibility rate with high-quality solutions, and outperformed classical BD approaches using simulated annealing.

Although our previous proof-of-concept work yielded promising results, extended evaluations have revealed several limitations that must be addressed to enhance the robustness of our framework. We generated additional random MILP instances following the structure of our original dataset and also created a new dataset with a different structure. These experiments exposed three primary issues: (1) an increased qubit count, (2) reduced solution quality and feasibility, and (3) convergence challenges. The higher qubit count arises from converting the MILP master problem into a QUBO model, which requires binary encoding of continuous variables and transforms constraints into quadratic penalty terms with associated penalty coefficients, and slack variables—a significant hurdle given current hardware constraints. Moreover, solution quality is highly sensitive to the tuning of penalty coefficients. In our earlier work, these penalties were manually set, risking underestimation which can prematurely halt the algorithm and yield infeasible or low-quality solutions, or overestimation which slows convergence and produces ineffective Benders cuts. Hence, precisely calibrated penalty values~\cite{ayodele2022penalty} are essential.  

On the other hand, evaluation on a new MILP dataset—where feasibility cuts naturally arise—highlighted another limitation: our original feasibility cut generation approach using CPLEX~\cite{IBMCPLEX2023} proved inadequate. Its reliance on extracting extreme rays via the Farkas certificate can fail when unboundedness is detected during the presolve phase~\cite{ibm_extreme_ray}, a common issue in small MILP instances. These findings clearly underscore the need for a more robust and versatile algorithm capable of handling larger and more diverse datasets. One final limitation we observed is the sequential generation of relatively weak Benders cuts, which further slows convergence.

Our contributions in this work address these limitations. First, we propose a novel approach for properly generating feasibility cuts, inspired by techniques from the L-shaped method~\cite{Lshaped1}, to overcome issues related to unreliable Farkas certificates. Second, we introduce improved strategies for optimizing qubit usage by tightening variable bounds and employing an exponential encoding method that reduces the need for slack variables. Third, in contrast to our previous work, where penalties were manually set, we develop constructive penalty tuning mechanisms to enhance solution quality. Finally, we present a multi-cut strategy that selects high-density cuts to accelerate algorithm convergence instead of relying on a sequential selection of single weak Benders cuts.

This paper is organized as follows. Section~\ref{sec:background} presents a brief mathematical background. Section~\ref{sec:feasibility_generator} describes our feasibility cuts generator. Section~\ref{sec:quit} details our approaches for optimizing qubit usage. Section~\ref{section:penalty} focuses on penalty tuning. Section~\ref{sec:enhancing_convergence} introduces our multi-cut strategy to enhance convergence. Performance evaluation is presented in Section~\ref{sec:results}. Finally, Section~\ref{sec:conclusion} concludes the paper.

\section{Brief Mathematical Background}
\label{sec:background}

In this section, we provide a brief mathematical background on classical Benders Decomposition and describe its extension to a hybrid classical–quantum framework via MILP-to-QUBO conversion.

\subsection{Principle of Classical Benders Decomposition}

Let \(A \in \mathbb{R}^{m_1 \times n}\), \(G \in \mathbb{R}^{m_1 \times p}\), and \(B \in \mathbb{R}^{m_2 \times n}\), and let the vectors \(c \in \mathbb{R}^n\), \(h \in \mathbb{R}^p\), \(b \in \mathbb{R}^{m_1}\), and \(b' \in \mathbb{R}^{m_2}\) be given. The original MILP (OP) is formulated as:
\begin{align}
\max_{x\in\{0,1\}^n,\, y\in\mathbb{R}_+^p}\quad & c^T x + h^T y \label{OP-related}\\
\text{s.t.}\quad & Ax+Gy\le b,\quad Bx\le b'. \label{y-variables-related}
\end{align}
BD splits the original MILP~\eqref{OP-related}-~\eqref{y-variables-related} into a MP over discrete variables, and a subproblem SP over continuous variables. For a fixed \(\hat{x}\) in the MP, the SP focuses on continuous variables. Its dual, SP-D, identifies extreme points and rays~\cite{schrijver1998theory}. By Minkowski’s theorem~\cite{bertsimas1997introduction}, these dual solutions form a finite basis for generating Benders’ cuts. Thus, the OP can be reformulated as:
\begin{align}
\label{master-problem-related}
\max_{x\in\{0,1\}^n,\ \phi\in\mathbb{R}} \quad & c^T x + \phi \\
\label{optimality-cut-related}
\text{s.t.}\quad & (b-Ax)^T\mu_o \ge \phi,\quad \forall\, \mu_o\in\mathcal{O}, \\
\label{feasibility-cut-related}
& (b-Ax)^T r_f \ge 0,\quad \forall\, r_f\in\mathcal{F}, \\
\label{x-cut-related}
& Bx \le b'.
\end{align}
Each extreme point \(\mu_o \in \mathcal{O}\) yields an optimality cut~\eqref{optimality-cut-related}, and each extreme ray \(r_f \in \mathcal{F}\) produces a feasibility cut~\eqref{feasibility-cut-related}. The algorithm begins with no cuts and iteratively solves the MP and subproblems, adding cuts until the subproblem objective meets or exceeds \(\phi\) (see~\cite{rahmaniani2017benders} for more details).

\subsection{From classical BD to HBD: MILP-to-QUBO Conversion}

In HBD, the MP is solved on a quantum processing unit (QPU) by converting the MILP formulation~\eqref{master-problem-related}-~\eqref{x-cut-related} into a QUBO model~\cite{glover2018tutorial}. For a detailed description of the QUBO-to-MILP conversion, the reader is referred to~\cite{naghmouchi2024mixed}. In the following, we summarize the key elements of the conversion that are used throughout this paper.

Starting from the formulation~\eqref{master-problem-related}–\eqref{x-cut-related}, the conversion proceeds as follows. In the MP objective~\eqref{master-problem-related}, the term \(c^T x\) is directly mapped to the quadratic form \(x^T \mathrm{diag}(c)x\) as the decision vector \(x\) is binary. The continuous variable \(\phi\) is transformed into a binary representation using the Hamiltonian
\begin{equation}
\label{eq:hp}
    H_\phi = \sum_{i=0}^{P-1} 2^i\, w_i \;+\; \sum_{j=1}^{D} 2^{-j}\, w_{P+j} \;-\; \sum_{k=1}^{N} 2^{k-1}\, w_{P+D+k},
\end{equation}
where  \(w_i\) are binary decision variables, and \(P\), \(D\), and \(N\) denote the numbers of bits allocated to the integer, fractional, and negative parts of \(\phi\), respectively. These parameters are determined by the upper and lower bounds of \(\phi\), denoted \(\phi_{\text{max}}\) and \(\phi_{\text{min}}\).

To encode the MP constraints in~\eqref{x-cut-related} into the QUBO, slack positive and continuous variables \(s_m\) are introduced so that \(Bx + s_m = b'\).
Each component \(s_m^k\) is binary-encoded and associated with a penalty coefficient \(\pi_1^k\). This results in the Hamiltonian
\begin{equation}
\label{master-qubo-constraints}
\nonumber
H_M = \sum_{k=1}^{m_2} \pi_1^k \,\bigl(B_k x + s_m^k - b'_k\bigr)^2.
\end{equation}
Similarly, each Benders cut—whether it is an optimality cut~\eqref{optimality-cut-related} or a feasibility cut~\eqref{feasibility-cut-related}—introduces an additional slack variable that is binary-encoded. The corresponding penalty Hamiltonians, \(H_O\) for optimality cuts and \(H_F\) for feasibility cuts, are then added.

The overall QUBO Hamiltonian is given by
\[
H_P = H_\phi + x^T \mathrm{diag}(c)x + H_M + H_O + H_F,
\]
which is minimized by the QPU.

In the next section, we introduce a feasibility constraint generator—an essential element not considered in our original PoC.

\section{Feasibility cut generator}
\label{sec:feasibility_generator}

Using a state-of-the-art solver to solve the SP is not always beneficial for generating feasibility cuts. For instance, the classical solver CPLEX~\cite{IBMCPLEX2023} relies on the Farkas certificate to return a dual unbounded direction by extracting extreme rays. However, if unboundedness is detected during the presolve phase, the extreme ray provided may be of poor quality~\cite{ibm_extreme_ray}, which is particularly problematic for small MILP instances—precisely the focus of our work. We propose here an alternative approach to reliably generate feasibility cuts. We draw inspiration from the L-shaped method commonly used in stochastic programming~\cite{Lshaped1}.

Our method proceeds as follows. Suppose that for a given solution \(\hat{x}\) of the MP, the original SP is infeasible. To address this infeasibility, we first reformulate the SP by adding one positive continuous slack variable to each constraint, thereby obtaining a modified subproblem \(SP2\):
\begin{align}
\label{SubFeas}
\nonumber \min_{\substack{y\in\mathbb{R}_+^p,\, s\in\mathbb{R}_+^{m_1}}} \quad & e^T s \\
\text{s.t.} \quad & A\hat{x}+Gy-s\le b.
\end{align}
with \(e^T = (1, \dots, 1)\). Here, \(SP2\) is always feasible. In the case that the original SP is feasible, the optimal solution of \(SP2\) will have \(s = 0\) and an objective value of zero; otherwise, a nonzero value of \(e^T s\) indicates the degree of infeasibility.

To exclude the current solution \(\hat{x}\) from the MP's feasible region in subsequent iterations, we must derive an appropriate feasibility cut. This is achieved by considering the dual of \(SP2\), denoted as \(D\text{-}SP2\):
\begin{align}
\label{dualsubfeas}
\max_{\mu\in\mathbb{R}_-^{m_1}} \quad & (b-A\hat{x})^T\mu \nonumber\\
\text{s.t.} \quad & G^T\mu \le 0,\quad -\mu_i \le 1,\quad i=1,\ldots,m_1.
\end{align}
Determining the feasibility cut is equivalent to achieving an objective value of zero in \(SP2\) (i.e., \(e^T s = 0\)), which, by the strong duality theorem, is equivalent to satisfying the inequality
$(b-A\hat{x})^T\mu \leq 0$, 
for some extreme point \(\mu\) of the dual feasible region. Consequently, we impose the following feasibility cut on the MP:
\begin{align}
(b-Ax)^T\mu &\le 0, \label{feasibility_cut}
\end{align}
thereby ensuring that any candidate solution \(x\) violating condition~\eqref{feasibility_cut}—including the current \(\hat{x}\)—is excluded from further consideration.

\section{Optimization of the number of qubits in the MILP-to-QUBO conversion}
\label{sec:quit}
In this section, we describe our strategies for optimizing the number of qubits required during the MILP-to-QUBO conversion process. In particular, we focus on two key aspects: (i) the encoding of the continuous master problem variable \(\phi\), and (ii) the conversion of constraints via either a slack variable-based or a slack-free (exponential) method.

\subsection{Qubits needed for continuous master problem variable \(\phi\)}
To encode the free continuous variable \(\phi\) in the master problem, we first tighten its bounds via linear programming (LP). Specifically, we define the lower and upper bounds as
\begin{equation}
\label{eq:lb}
lb = \min_{\substack{x \in [0,1]^n \\ y \in \mathbb{R}_+^m}} \{ h^T y \mid Ax + Gy \leq b,\; Bx \leq b' \},
\end{equation}
\begin{equation}
\label{eq:ub}
ub = \max_{\substack{x \in [0,1]^n \\ y \in \mathbb{R}_+^m}} \{ h^T y \mid Ax + Gy \leq b,\; Bx \leq b' \}.
\end{equation}
These bounds determine the qubit allocations for \(\phi\). We have that
\[
\begin{aligned}
P &= \lfloor \log_2(\lfloor ub \rfloor) \rfloor + 1, 
D &= \lfloor \log_2(1/\epsilon) \rfloor + 1, 
N &= \lfloor \log_2(|lb|) \rfloor + 1.
\end{aligned}
\]
Here, \(\epsilon\) is the desired precision. Note that if \(ub \le 0\), no bits (qubits) are needed for the positive part; if \(lb \ge 0\), no bits (qubits) are required for the negative part. This LP-based tightening technique was mentioned in~\cite{naghmouchi2024mixed} but has not been previously implemented, and weak bounds were instead considered.

\subsection{Qubits needed for slack variables encoding}
Consider an optimality cut associated with an extreme point \(\mu_o\), the corresponding positive continuous slack variable is $s_o = b^T\mu_o - (Ax)^T\mu_o - \phi.
$ The upper bound for \(s_o\) is determined by solving the LP
\begin{equation}
\label{eq:so_lp}
\max_{\substack{x\in[0,1]^n,\; y\in\mathbb{R}_+^m \; \phi \in \mathbb{R} \\ Ax+Gy\le b,\; Bx\le b'}} \Bigl\{ b^T\mu_o - (Ax)^T\mu_o - \phi \Bigr\}.
\end{equation}

Since \(s_o \geq 0\) by construction, no qubits are allocated for its negative part. This approach is similarly applied to slack variables used in feasibility cuts.

\subsection{Exponential method: slack-variable-free conversion}
Alternatively, we propose an exponential method that eliminates the need for slack variables as suggested in ~\cite{expo}. Given a constraint \(h(x) \le a\), define 
$g(x)=h(x)-a, f(x)=e^{g(x)}-1$. When \(g(x)\le0\), \(f(x)\) remains in \([-1,0]\); when \(g(x)>0\), \(f(x)\) increases rapidly. Approximating \(f(x)\) via its second-order Taylor expansion yields $f(x)\approx g(x)+\frac{1}{2}g(x)^2$.

The constraint is then incorporated into the objective as a penalty term: $\pi \Bigl(g(x)+\frac{1}{2}g(x)^2\Bigr)$,
where \(\pi\) is a penalty coefficient that must be carefully tuned. This formulation eliminates the introduction of slack variables, thus keeping the qubit count constant.

Note that this process may introduce numerical imprecisions due to the approximate encoding of the constraints. However, as our experimental results in Section~\ref{sec:results} demonstrate, the numerical outcomes are satisfactory provided that \(\pi\) is properly tuned. In the next section, we present a constructive method for tuning the penalty parameters during the MILP-to-QUBO conversion.

\section{Constructive Method for Penalty Tuning}
\label{section:penalty}

Given a MILP formulation of the OP, the penalty tuning problem involves determining penalty coefficients for the various terms in the QUBO such that (i) every solution that is feasible for the OP achieves a lower QUBO objective value than any infeasible solution, and (ii) the relative ordering of feasible solutions is preserved. In other words, if a feasible solution \(x_1\) is better than \(x_2\) in the OP, then the corresponding QUBO objective values should reflect the same ranking. It is worth noting that this penalty tuning problem can be even more challenging than solving the MILP itself~\cite{ayodele2022penalty}. In fact, exactly solving the penalty tuning problem may require advanced and time-consuming techniques, such as iterative cut generation, and remains an open research topic.

In our previous work~\cite{naghmouchi2024mixed}, penalty coefficients were set manually based on problem-specific data. Here, we propose a constructive method that systematically derives the penalties from an upper bound \(Ub\) on the master problem objective \(c^T x + \phi\). Specifically, we define $Ub = \left|\sum c\right| + |\phi_{\text{max}}| + 1$,
where \(\phi_{\text{max}}\) is obtained from~\eqref{eq:ub}.

Our approach distinguishes the penalty parameters for different parts of the QUBO formulation. Let \(\pi_{\text{obj}_x}\) denote the penalty for the \(x\)-dependent component of the objective~\eqref{master-problem-related}, \(\pi_{\text{obj}_\phi}\) the penalty for the \(\phi\)-dependent component in~\eqref{eq:hp}, \(\pi_{\text{obj\_cut}}\) the penalty associated with the Benders cut Hamiltoians $H_O$ and $H_F$, and \(\pi_{\text{cons}_{MP}}\) the penalty for enforcing the master problem constraints in~\eqref{master-qubo-constraints}. In our formulation, these penalties are set as
\[
\pi_{\text{obj}_x}=3\,Ub, \pi_{\text{obj}_\phi}=Ub, \pi_{\text{obj\_cut}}=Ub, \pi_{\text{cons}_{MP}}=Ub^2.
\]

By using \(Ub\) as a baseline scaling factor derived from the MILP instance, our constructive method assigns higher weights to terms that must be strictly enforced (such as the master constraints) while allocating moderate weights to the objective components and Benders cuts.

\section{Multi-cut approach for enhancing algorithm convergence}
\label{sec:enhancing_convergence}

Standard Benders Decomposition is known to suffer from slow convergence. To mitigate this, we generate multiple cuts per iteration rather than a single cut. However, many candidate cuts exhibit low density—that is, most coefficients corresponding to MP decision variables are zero or nearly zero—thereby limiting their effect in tightening the MP. In contrast, high-density cuts more effectively restrict the solution space and accelerate convergence~\cite{PATERAKIS2023108161}.

Our approach retains the \(k\) best solutions from the MP, for which \(k\) candidate cuts are generated. To avoid an excessive number of cuts—which would increase the qubit overhead when encoding slack variables—only a subset of these candidate cuts is selected. Cut selection is based on the density of the cuts. To this end, we construct a density matrix \(D \in \{0,1\}^{k \times |\Gamma|}\), where \(\Gamma\) denotes the set of indices for the MP decision variables. In this matrix, each entry \(D_{k,n}\) is set to 1 if the coefficient of decision variable \(x_n\) in candidate cut \(k\) is nonzero, and 0 otherwise. This matrix encapsulates the density information of the candidate cuts. 

We then solve an unweighted Maximum Coverage Problem (MCP)~\cite{takabe2018typical} to select the subset of candidate cuts that collectively cover as many MP decision variables as possible, subject to an upper bound \(M\) (with \(M \leq k\)) on the number of cuts. An ILP formulation of the MCP can be found in~\cite{PATERAKIS2023108161}. Currently, we solve the MCP exactly using a classical solver, as our experimental instances are relatively small. In the future, as quantum technology advances and instance sizes grow, heuristic approaches for solving the MCP may become more appropriate.

\begin{figure*}
    \centering
    \begin{subfigure}[t]{0.3\textwidth}
        \centering
        \raisebox{0cm}{
            \includegraphics[width=0.9\linewidth]{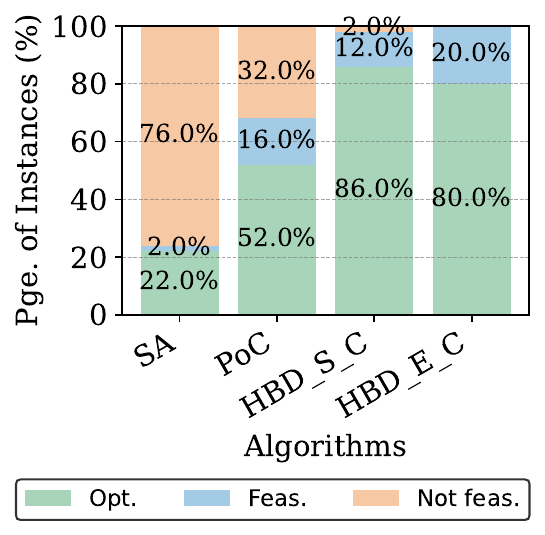} 
        }
        \caption{Solution feasibility}
        \label{fig:feasibility}
    \end{subfigure}
    \hspace{0.4cm} 
    \begin{subfigure}[t]{0.3\textwidth}
        \centering
        \raisebox{0.5cm}{ 
            \includegraphics[width=0.9\linewidth]{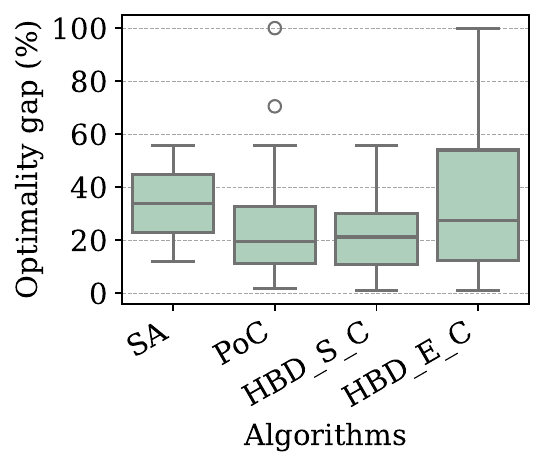}
        }
        \caption{Optimality gap}
        \label{fig:gap}
    \end{subfigure}
    \hspace{0.4cm} 
    \begin{subfigure}[t]{0.3\textwidth}
        \centering
        \raisebox{0.5cm}{
            \includegraphics[width=0.9\linewidth]{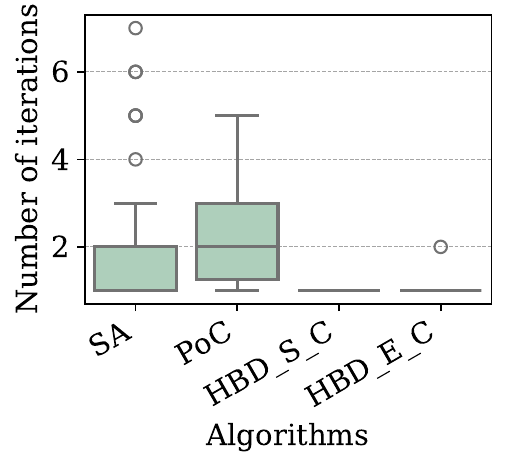}
        }
        \caption{Convergence (number of iterations)}
        \label{fig:iteration}
    \end{subfigure}
    \vspace{-0.1cm} 
    \caption{Performance comparison with the MILPs data set from~\cite{naghmouchi2024mixed}.}
    \label{fig:global}
\end{figure*}
\begin{figure*}
    \centering
    \begin{subfigure}[t]{0.3\textwidth}
        \centering
        \raisebox{0cm}{
            \includegraphics[width=0.9\linewidth]{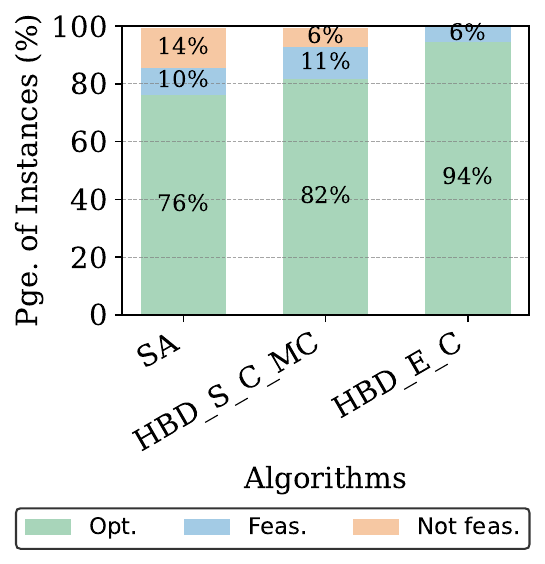}
        }
        \caption{Solution feasibility}
        \label{fig:feasibility_MILP2}
    \end{subfigure}
    \hspace{0.4cm} 
    \begin{subfigure}[t]{0.3\textwidth}
        \centering
        \raisebox{0.5cm}{
            \includegraphics[width=0.9\linewidth]{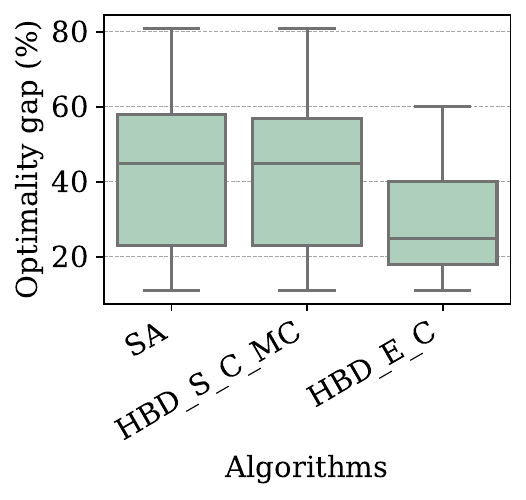}
        }
        \caption{Optimality gap}
        \label{fig:gap_MILP2}
    \end{subfigure}
    \hspace{0.4cm} 
    \begin{subfigure}[t]{0.3\textwidth}
        \centering
        \raisebox{0.5cm}{
            \includegraphics[width=0.9\linewidth]{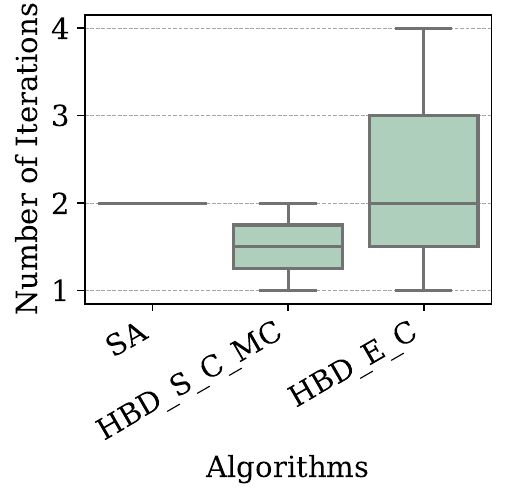}
        }
        \caption{Convergence (number of iterations)}
        \label{fig:iterations_MILP2}
    \end{subfigure}
    \vspace{-0.1cm} 
    \caption{Performance evaluation on the generic MILPs data set.}
    \label{fig:global_MILP2}
\end{figure*}

\section{Numerical results}
\label{sec:results}

In this section, we present a series of numerical experiments to assess the impact of the reinforcements proposed in this paper on the HBD framework. Specifically, we compare the enhanced framework against the original PoC algorithm from~\cite{naghmouchi2024mixed}. In addition, we evaluate these reinforcements on a more generic dataset than the one previously used, allowing us to examine their effect on instances where feasibility cuts naturally arise. Note that in the first dataset, the MILP instances' structure and size did not allow for feasibility cuts.

\subsection{Instances description and experimental setting}

We conduct experiments on two datasets. The first is identical to that used in our original BD framework~\cite{naghmouchi2024mixed}, enabling direct comparison with our enhanced algorithm, while the second comprises a broader set of MILPs that require at least one feasibility cut, thereby testing the algorithm’s versatility.

The generic MILPs in the second dataset is composed of 125 instances constructed as follows. The MILP comprises 2–5 binary variables \(x\), and 2–10 continuous non-negative variables \(y\). Matrices \(A\) and vector \(b\) contain positive integers in \([0, 10]\); matrix \(G\) has dimensions between 5 and 14 with entries in \([-5, 5]\); matrix \(B\) consists entirely of ones with \(b'\) less than 5; and vectors \(c\) and \(h\) contain positive integers in \([0, 10]\).

Our implementation is in Python. The quantum algorithm is simulated using Pulser~\cite{PulserDocs} (neutral atom devices with a 12-qubit limit) and the variational parameters are tuned using scikit-optimize~\cite{scikit-optimizeDocs}. We use CPLEX~\cite{IBMCPLEX2023} to solve to optimality (i) the OP~\eqref{OP-related}--\eqref{y-variables-related} to determine optimality gaps, (ii) the LPs~\eqref{eq:lb} and~\eqref{eq:ub} to compute bounds for the continuous variables, and (iii) the ILP formulation of the MCP for multi-cut selection. All experiments are executed on an AMD EPYC 7282 16-Core processor (64-bit mode, 2800 MHz).

Throughout this section, we use the following notations: SA denotes the Simulated Annealing algorithm; PoC refers to the Proof-of-Concept BD algorithm~\cite{naghmouchi2024mixed}; and HBD variant denotes an enhanced hybrid BD algorithm, characterized by its MILP-to-QUBO conversion (E for Exponential method, S for Slack method), penalty tuning (C for Constructive, M for Manual), and optionally the multicut method (MC). For example, \texttt{HBD\_E\_C} denotes a variant using the Exponential conversion method and constructive penalty tuning, without multicut.

The performance metrics we evaluate are: \emph{feasibility Rate} (percentage of solutions satisfying all constraints); \emph{optimality Rate} (percentage of optimal solutions); \emph{optimality Gap}: $\frac{\mathrm{opt} - \mathrm{obj(A)}}{\mathrm{opt}}$,
where \(\mathrm{obj(A)}\) is the objective value achieved by the considered algorithm (enhanced HBD or SA) and \(\mathrm{opt}\) is the optimal objective value; and \emph{BD Iterations} (the number of iterations performed in the Benders Decomposition process).

\subsection{Performance Evaluation}  
An extensive experimental study was conducted to determine the optimal combination of algorithmic enhancements for each dataset. The results presented in the following sections are based on the configurations that yielded the best performance.

\subsubsection{Comparison with the PoC Algorithm}

Figure~\ref{fig:feasibility} illustrates how our algorithmic reinforcements positively affect feasibility by categorizing outcomes as optimal (Opt.), feasible (Feas.), or not feasible (Not feas.). The standard SA algorithm yields 22\% optimal and only 2\% feasible solutions, leaving 76\% of the instances infeasible. By contrast, our PoC enhances optimality to 52\%, with 16\% optimal solutions, while reducing infeasibility to 32\%. These improvements align with our earlier findings in~\cite{naghmouchi2024mixed}. Among the two best variants, HBD\_S\_C (Slack) achieves 86\% optimal and 12\% feasible-but-not-optimal solutions (with 2\% infeasibility), while HBD\_E\_C (Exponential) attains 80\% optimal and 20\% feasible solutions with 0\% infeasibility.
In other words, one may accept a small fraction of infeasibility to gain a slight advantage in optimal solutions with HBD\_S\_C, or choose HBD\_E\_C to guarantee full feasibility at a marginal reduction in the share of optimally solved instances.

Turning to solution quality, Figure~\ref{fig:gap} reports the optimality gap for all feasible solutions. SA starts with a relatively high median gap (around 35--40\%), while PoC improves it to about 20\%. HBD\_S\_C maintains a similar median gap to PoC, whereas HBD\_E\_C exhibits a higher gap (the highest among all tested methods). Thus, if one opts to remove infeasibility entirely by employing HBD\_E\_C, the trade-off is a lower quality of feasible solutions on average. Finally, Figure~\ref{fig:iteration} reveals the computational advantage of our improved HBD approaches. SA typically requires a median of two iterations to converge, with outliers occasionally reaching seven. With PoC, we observe a median of two iterations but still find outliers beyond four or five. In contrast, both HBD\_S\_C and HBD\_E\_C generally converge in one iteration, with minimal dispersion. This result indicates that the reinforcement strategies introduced—generating enhanced Benders’ cuts—substantially reduce the algorithmic effort needed to reach a final solution.

Overall, Figures~\ref{fig:feasibility}, \ref{fig:gap}, and \ref{fig:iteration} confirm that our enhancements reduce infeasibility, improve optimality, and lower iteration counts compared to the PoC, thereby validating our approach.

\subsubsection{Performance evaluation on generic MILPs data set}
\label{sec:milp2}
From Figure~\ref{fig:feasibility_MILP2}, SA yields 76\% optimal solutions, 10\% feasible-but-not-optimal, and 14\% infeasible. With HBD\_S\_C\_MC (Slack method combined with constructive penalties and multi-cuts), these percentages improve to 82\% optimal and 11\% feasible, leaving only 6\% infeasible. HBD\_E\_C pushes optimal solutions even higher (94\%), at the cost of 6\% non-optimal feasibility, but no unfeasible outcomes. 

Regarding solution quality (Figure~\ref{fig:gap_MILP2}), we observe that the median optimality gap for SA is around 40\%, with values reaching up to 80\%. HBD\_S\_C\_MC exhibits a similar median gap, whereas HBD\_E\_C reduces it to roughly 25\%, reflecting an overall improvement in solution quality. Finally, Figure~\ref{fig:iterations_MILP2} highlights iteration counts. SA typically converges in about two iterations, while HBD\_S\_C\_MC exhibits a lower median— two iterations—and minimal dispersion. By contrast, HBD\_E\_C is more variable: its median is two, but with whiskers extending up to four iterations. This observation suggests the exponential method can eliminate infeasibility and improve the quality of feasible solutions, though it may require slightly higher computational effort.

These results demonstrate that our reinforcement strategies—especially the use of multiple cuts—yield significant improvements in feasibility, solution quality, and computational efficiency on generic MILPs. While HBD\_E\_C achieves a higher proportion of optimal solutions, HBD\_S\_C\_MC offers a competitive trade-off by maintaining robust solution quality relative to SA. Moreover, unlike the PoC data set, introducing multiple cuts proves beneficial.

\section{Conclusion and perspectives}
\label{sec:conclusion}
In this paper, we have presented key enhancements to our quantum neutral-atom-assisted Benders Decomposition framework for solving MILPs. Our contributions include a robust feasibility cut generator; an optimized MILP-to-QUBO conversion that reduces qubit requirements via tightened variable bounds and an exponential encoding method; a constructive penalty tuning mechanism; and finally a multi-cut strategy to accelerate convergence. Experimental results confirm notable improvements in feasibility, solution quality, and convergence compared to our previous PoC as well as classical methods. Future work will focus on further tightening continuous variable bounds to minimize qubit usage, and advancing penalty tuning via cutting planes and/or machine learning—promising further enhancements in the performance of the hybrid Benders Decomposition assisted with neutral atoms.

\section*{Acknowlegments}
We thank Boris Detienne for insightful discussions regarding feasibility cuts. 

\bibliographystyle{IEEEtran}
\bibliography{bibliography}

\end{document}